\documentclass[a4paper,fleqn,usenatbib]{mnras}

\usepackage{newtxtext,newtxmath}

\usepackage[T1]{fontenc}
\usepackage{ae,aecompl}

\usepackage{graphicx}	
\usepackage{amsmath}	
\usepackage{amssymb}	

\title[The first sub-70 minute non-interacting WD-BD system]{The first sub-70 minute non-interacting WD-BD system: EPIC212235321}
\author[S. L. Casewell]{S. L. Casewell$^{1}$ \thanks{E-mail: slc25@le.ac.uk},  I. P. Braker$^{1}$, S. G. Parsons$^{2}$, J. J. Hermes$^{3}$ \thanks{Hubble Fellow}, M.R. Burleigh$^{1}$, \newauthor C. Belardi$^{1}$, A. Chaushev$^{1}$,  N. L. Finch$^{1}$, M. Roy$^{1}$, S. P. Littlefair$^{2}$, M. Goad$^{1}$ \newauthor
and E. Dennihy$^{3}$\\
$^{1}$Department of Physics and Astronomy, Leicester Institute of Space and Earth Observation, University of Leicester,\\
University Road, Leicester LE1 7RH, UK \\
$^{2}$Department of Physics and Astronomy, University of Sheffield, Sheffield, S3 7RH, UK\\
$^{3}$Department of Physics and Astronomy, University of North Carolina, Chapel Hill, NC 27599-3255, USA}
\date{Accepted XXX. Received YYY; in original form ZZZ}

\pubyear{2017}
\begin{document}
\label{firstpage}
\pagerange{\pageref{firstpage}--\pageref{lastpage}}
\maketitle

\begin{abstract}
We present the discovery of the shortest-period, non-interacting, white dwarf-brown dwarf post-common-envelope binary known. The K2 light curve shows the system, EPIC 21223532 has a period of 68.2 min and is not eclipsing, but does show a large reflection effect due to the irradiation of the brown dwarf by the white dwarf primary. Spectra show hydrogen, magnesium and calcium emission features from the brown dwarf's irradiated hemisphere, and the mass indicates the spectral type is likely to be L3. Despite having a period substantially lower than the cataclysmic variable period minimum, this system is likely a pre-cataclysmic binary, recently emerged from the common-envelope. These systems are rare, but provide limits on the lowest mass object that can survive common envelope evolution, and information about the evolution of white dwarf progenitors, and post-common envelope evolution.  

\end{abstract}



\section{Introduction}
To date, there are only a handful of known close, detached, post-common envelope systems containing a white dwarf with a brown dwarf companion. This is not surprising, as brown dwarf companions to main sequence stars within 3\,AU are rare compared to planetary or stellar companions to main sequence stars \citep{grether06, metchev04}.  For instance, the SuperWASP survey \citep{pollacco06} has discovered over 120 confirmed extrasolar planets, but only one brown dwarf, WASP-30b \citep{anderson11}, within the same parameter space, although this may be due to selective follow-up.  Brown dwarf companions to white dwarf stars are even rarer with only 0.5\% of white dwarfs having a brown dwarf companion \citep{steele11}. 
Despite many candidate systems being discovered from all-sky surveys (e.g. \citealt{girven11,debes11}), only eight
post-common envelope systems have been confirmed: GD1400 (WD+L6, P=9.98\,hrs; \citealt{farihi04, dobbie05, burleigh11}), WD0137-349 (WD+L6-L8, P=116\,min; \citealt{maxted06, burleigh06}), WD0837+185 (WD+T8, P=4.2\,hrs; \citealt{casewell12}), NLTT5306 (WD+L4-L7, P=101.88\,min; \citealt{steele13}), SDSS J155720.77+091624.6 (WD+L3-L5, P=2.27\,hrs); \citealt{farihi17}, SDSS J1205-0242 (WD+L0, P=71.2\,min; \citealt{parsons17,rappaport17}), SDSS J1231+0041 (WD+M/L, P=72.5\,min; \citealt{parsons17}) and SDSS J141126.20+200911.1, (WD+T5, P=121.73\,min; \citealt{beuermann13, littlefair14}). All of these systems have survived a phase of common-envelope evolution, resulting in the close binary system. These systems are all detached, and the brown dwarf is likely tidally-locked. Eventually, they will become cataclysmic variables, such as SDSS1433+1011 in which the substellar donor was recently detected \citep{hernandez16}.

Along with CVs containing a brown dwarf, these few systems represent the lowest mass objects that can survive common envelope evolution, as well as informing us about the mass loss processes as the giant evolves (e.g. \citet{rappaport17,casewell12}). In the case of SDSS J155720.77+091624.6, there is not only a brown dwarf that has survived the common envelope evolution, but also a dust disk that may represent the remnants of a planetary system \citep{farihi17}.    Additionally, these systems can provide information on how substellar objects respond to heating. Spectra of WD0137-349B show emission lines consistent with the cloud composition of an L dwarf, suggesting that these objects can be used as directly detected proxies for exoplanets.

Here we present the discovery of the shortest-period, non-interacting, white dwarf-brown dwarf post-common-envelope binary known to date.

\section{Kepler observations}
\begin{figure}
 \begin{center}
 \includegraphics[width=\columnwidth]{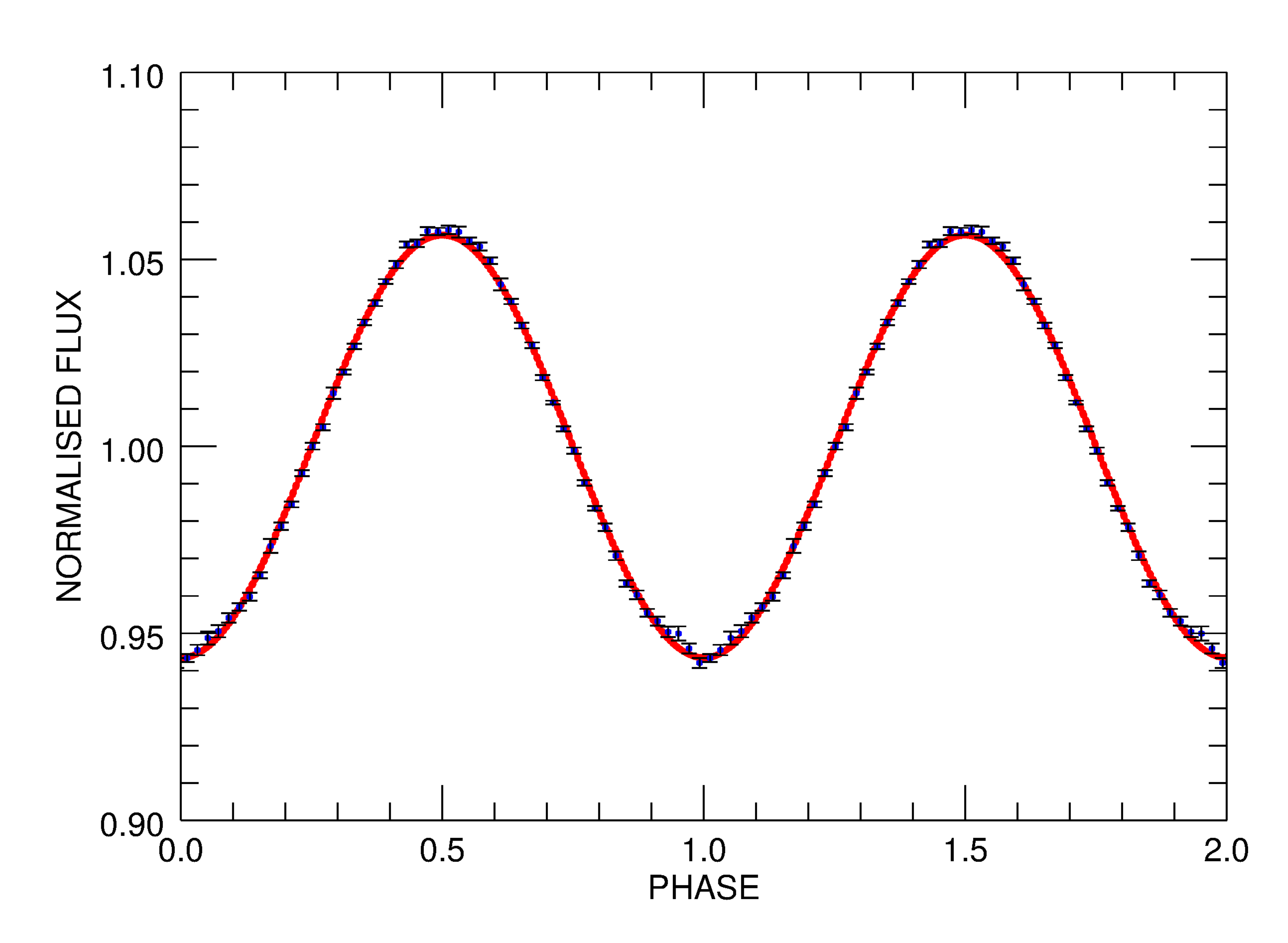}
    \caption{Phase folded K2 long cadence data lightcurve for EPIC21223535321 binned over 50 equally spaced bins. The red line shows the sinusoidal fit to the lightcurve.}
     \label{fig:example_figure}
     \end{center}
 \end{figure}

The Kepler 2 (K2) mission \citep{howell14} has observed hundreds of spectroscopically and photometrically identified white dwarfs. EPIC212235321 was identified as a photometric candidate white dwarf from the SuperCosmos Sky Survey \citep{hambly01_1,hambly01_2,hambly01_3} and was proposed as a target by PI Kilic in Campaign 3 field (RA 22:26:40, DEC -11:05:48). See Table \ref{tab:info} for the photometric parameters. EPIC212235321 was observed by K2 for $\approx69$ days between 2014 Nov 15 and 2015 Jan 23 in long cadence mode. Our analysis used the K2 Guest Observer Office \citep{vancleve16} lightcurve obtained through the Mikulski Archive for Space Telescopes (MAST). The light curve was normalised and flagged points (such as those affected by cosmic ray hits) were removed resulting in a lightcurve with 3151 data points. We searched for periodicity in the  lightcurve using Lomb-Scargle \citep{Lomb76,Scargle82} and discrete fourier transform periodograms in \textsc{idl} and the periodogram software packages \textsc{Vartools} \citep{Hartman16} and Period04 \citep{lenz05}. These identified a most likely period of $0.047369569 \pm 0.000000056$ days ($\approx$68.21 minutes) with period uncertainties determined using the bootstrap resampling (with replacement) methodology detailed in \citealt{lawrie13_1}; with a normalised amplitude of $0.05650 \pm 0.00009$. A Monte Carlo significance test was undertaken on the period as outlined in \citealt{cumming99, lawrie13_2} yielding a False Alarm Probability of $<0.001$. The lightcurve is shown in Figure \ref{fig:example_figure}. A sine curve was fitted to the data using MPFIT \citep{markwardt09} as implemented by \citealt{lawrie13_2} giving a combined ephemeris of BJD = (2457011.653656 $\pm$ 0.000013) + (0.047369569 $\pm$ 0.000000056) * E.

\begin{table}
	\centering
	\caption{Photometry for EPIC212235321}
	\label{tab:info}
	\begin{tabular}{ccc} 
    \hline
	Property	&	Value		&info\\
	\hline
	Ra&22:03:40.61 &\\
    Dec&-12:15:10.8 &\\
    K2 &17.6& K2\\
    fuv&16.573$\pm$0.030&\textit{Galex}\\
    nuv&16.866$\pm$0.022&\textit{Galex}\\
    $u$&17.122$\pm$0.007&VST ATLAS \\
    $g$&17.386$\pm$0.004&VST ATLAS \\
    $r$&17.649$\pm$0.004&VST ATLAS \\
    $i$&18.104$\pm$0.010&VST ATLAS \\
    $z$&18.299$\pm$0.021&VST ATLAS \\
    $Y$&17.542$\pm$0.017 &VISTA VHS  \\
    $J$&17.512$\pm$0.027  &VISTA VHS  \\
    $K_s$&17.594$\pm$0.130&VISTA VHS\\
\hline
\end{tabular}
\end{table}
Due to the large number of points in the K2 lightcurve (3151), this fitting produced a very precise period which is still more precise than our follow-up high cadence data, taken a year later.

\section{Optical Photometry}
We observed EPIC212235321 using the Sutherland High-speed Optical Camera (SHOC; \citealt{coppejans}) camera on the SAAO 1.0~m telescope. The first observations were conducted in white light on the night of the 11th of July 2016, with subsequent observations conducted again on the 24th to 27th of November 2016 in the $g$ and $i$ band filters respectively. Observing conditions were affected by some scattered high cloud during the beginning of the night on the 11th of July which may have affected the white light observations. 

The data were reduced using the standard procedure with sky flats and bias frames taken during the observing run. We used the \textsc{starlink} package \textsc{autophotom} to perform the photometry of the target and comparison stars. The aperture was fixed for the data and was set to be two times the mean seeing (full width at half-maximum; \citet{Naylor98}). This aperture size limited the impact of the background noise in the aperture. The sky background level was determined using the clipped mean of the pixel value in an annulus around the stars and the measurement errors were estimated from the sky variance. To remove atmospheric fluctuations, the light curve was divided by the light curve of one of the comparison stars. 

\begin{table*}
	\centering
	\caption{Observations taken at the SAAO of EPIC212235321. The T$_0$ determined in each waveband is also given. The $i$ band data were combined to calculate one value for the waveband.}
	\label{tab:obs}
	\begin{tabular}{ccccc} 
	Date	&	filter		&Exp time (s) & Total time (hours)& T$_0$ (HJD)\\
	\hline
	20160711& white light&5&4.92&2457581.60276 $\pm$ 0.00013\\
    20161124&$i$&60&0.85&-\\
    20161125&$i$&30&1.36&-\\
    20161126&$g$&30&2.01&2457581.60225 $\pm$ 0.00085\\
    20161127&$i$&30&1.64&2457581.60371 $\pm$ 0.00032\\
\hline
\end{tabular}
\end{table*}
\begin{figure}
	\begin{center}
	\includegraphics[width=\columnwidth]{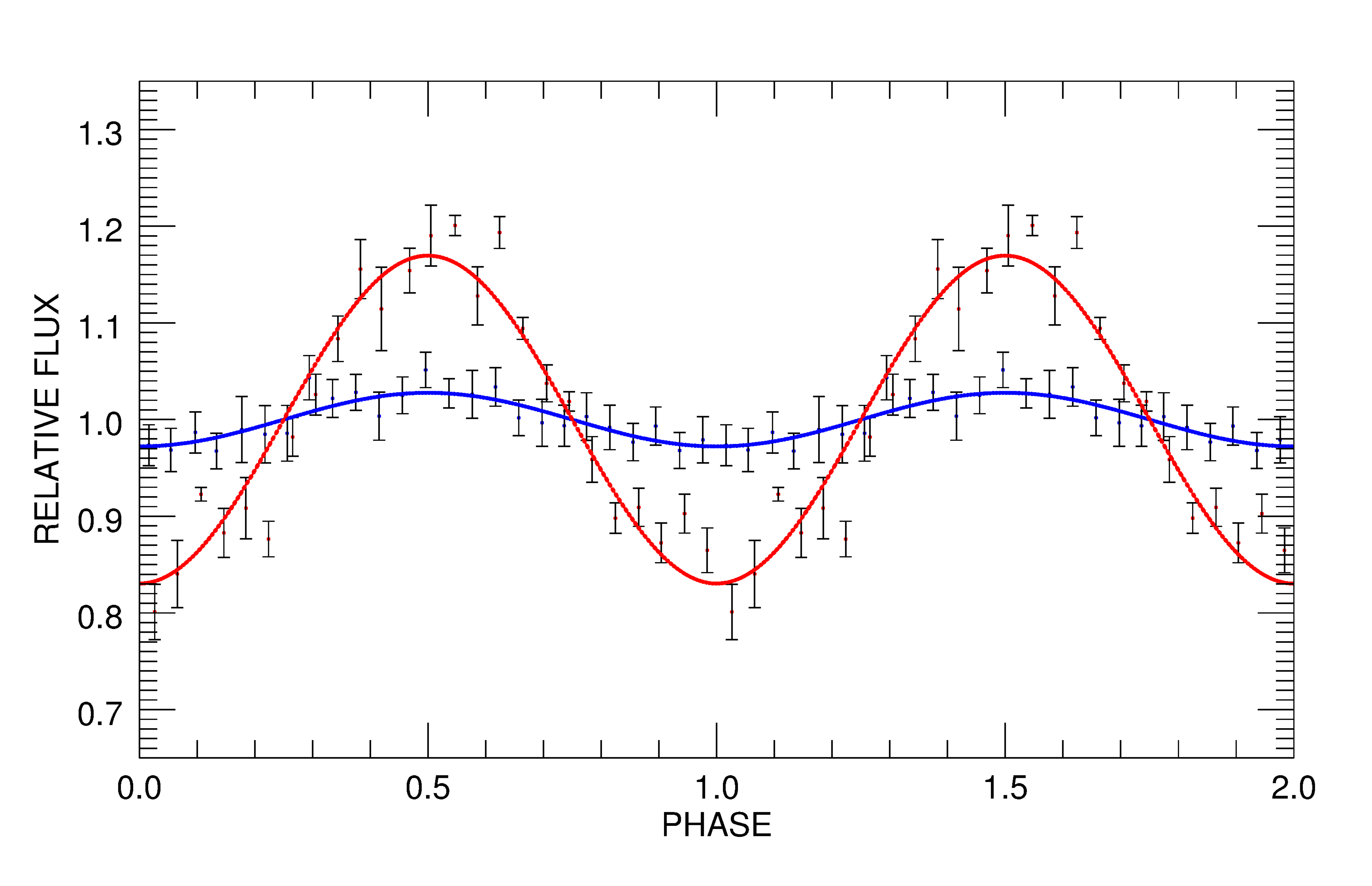}
    \caption{$g$ (blue) and $i$ (red) band lightcurves for EPIC21223535321. The semi-amplitudes were 0.028$\pm$0.003 mags in the $g$ band and  0.170$\pm$0.008 in the $i$ band. The photometry has been binned into 25 equal bins. The lightcurves have been duplicated over two periods for display purposes.}
    \label{lc}
\end{center}
\end{figure}
The $g$ and $i$ band lightcurves in Figure \ref{lc}, folded on the 68.21 minute period, show no eclipses, which would have been detected with the cadence of the observations. If present, these should have been identified within the K2 data, however, there is a chance that the long cadence of the data means a short eclipse was missed. The lightcurves have been fitted with a sine curve, and have slightly different T$_{0}$, indicative of a phase shift between the two bands, similar to that seen in WD0137-349 by \citet{casewell15}. The difference is only 2.1 minutes, however, this is slightly less than twice the maximum error derived for the T$_{0}$. The semi-amplitudes were 0.028$\pm$0.003 mags in the $g$ band and  0.170$\pm$0.008 in the $i$ band.  The K2 band (4200 \AA - 9000 \AA) semi-amplitude is 0.05650$\pm$0.00009, indicating that the reflection effect due to the irradiation of the brown dwarf is increasing  as we move to longer wavelengths. There is a discrepancy of $\approx$52s between the $i$ band and K2 T$_{0}$ derived from the $i$ band filter when compared to the K2 pass band, however this is is within a 1 sigma error in the period over the $\sim$12,000 cycles between the two T$_{0}$ values.

\section{ISIS spectroscopy}
To confirm the nature of the primary star, we observed  EPIC212235321 on the night of 20150909 with ISIS  on the William Herschel Telescope on La Palma. We used the 1" slit with the R600B grating in the blue arm and the GG495 filter with the R600R grating in the red arm. We took four exposures of 900~s, at airmass of $\sim$1.5 and seeing between 0.8" and 1.5". We  also observed arc lamps and flux standards to calibrate the data. The data were reduced using \textsc{iraf} routines for long slit spectroscopy  as in \citet{dobbie06}.

The spectra confirm that the target is indeed a white dwarf. We  determined the effective temperature (T$_{\rm eff}$)  and surface gravity (log g) of the white dwarf, from the Balmer absorption lines (omitting H$\alpha$ and  H$\beta$ as they are affected by emission features) by comparing to the predictions of white dwarf model atmospheres using the spectral fitting program FITSB2 (v2.24; \citealt{napiwotzki04}).

\begin{figure}
\begin{center}
	\includegraphics[trim=5cm 3cm 0cm 0cm,clip=yes,
    scale=0.5]{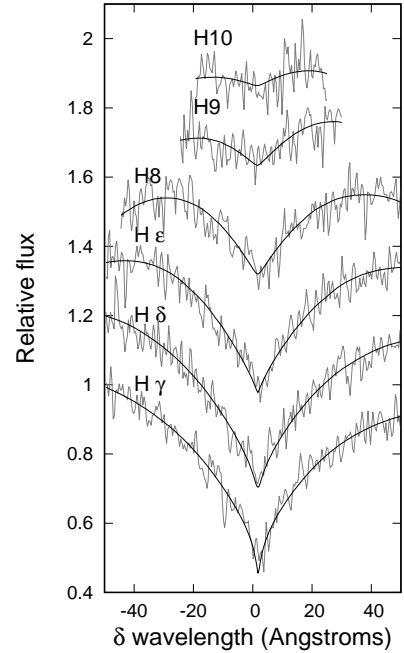}
    \caption{ISIS spectrum of EPIC212235321 with the best fitting model T$_{\rm eff}$ = 24490$\pm$150 K, log g = 7.63$\pm$0.02 overplotted. H$\alpha$ and H$\beta$ have been omitted as they are the most affected by the emission features.}
    \label{teff}
    \end{center}
\end{figure}

We generated pure hydrogen DA models with mixing length ML2/$\alpha$=0.8 \citep{koester10} and Balmer/Lyman lines calculated with the modified Stark broadening profiles of \citet{tremblay09}. We then generated a model grid from 6000 to 80000~K in 1000~K steps with log g ranging from 6.5 to 9.5 in 0.25~dex steps.  We used FITSB2 to fit our grid of model spectra to the six Balmer absorption lines ranging from H$\delta$ to H10 in each exposure. In addition, points in the observed data lying more than 3$\sigma$ from the model were clipped from subsequent iterations of the fitting process.  The  best fit values were T$_{\rm eff}$ = 24490$\pm$150~K, log g = 7.63$\pm$0.02 (Table \ref{rv_tab}, Figure \ref{teff}). 
%
%

\section{SOAR spectra}

We obtained 20 higher signal-to-noise  spectra using the Goodman spectrograph on the 4.1 m SOAR telescope \citep{clemens04}. 
These spectra, had 480 s exposures taken consecutively on 2016 August 15, cover H$\alpha$ using the R1200 grating, and range from roughly 6000  - 7300 {\AA} with a dispersion of roughly 0.3 {\AA} per pixel. We used a 1" slit, and the seeing was 0.9". The data were reduced in the same manner as the WHT spectra, with additional wavelength calibration performed using the skylines present in the spectra.

The SOAR spectra were normalised to 1 and phase-binned on the ephemeris using Tom
Marsh's \textsc{molly}\footnote{http://deneb.astro.warwick.ac.uk/phsaap/software/molly/html/INDEX.html} software.  There is clearly an emission feature in the centre of the H$\alpha$ line that moves in antiphase with the H$\alpha$ absorption features from the white dwarf. To model these absorption and emission features  we used the technique in \citet{parsons17}, using three Gaussian profiles to model the absorption from the white dwarf that change position according to $\gamma_1 + K_1 \sin(2\pi\phi)$, where $\phi$ is the orbital phase, and two Gaussian components to model the H$\alpha$ emission from the companion star that change position according to  $\gamma_{em} + K_{em} \sin(2\pi\phi)$, and vary in strength according to ($1-\cos\phi)/2$. We also allowed an offset to be fit to take into account the error on the period.  The errors on the radial velocity parameters were determined as in \citet{casewell15}, with 1 kms$^{-1}$ added in quadrature in order to achieve a reduced chi-squared with $\chi_{\nu}^2\sim$~1. The final velocities were K$_1$=42$\pm$4 km~s$^{-1}$ and K$_{em}$=305$\pm$7 km~s$^{-1}$ with $\gamma_1$=30$\pm$2 km~s$^{-1}$ and $\gamma_{em}$=-4$\pm$6 km~s$^{-1}$. There is a notable difference in the gamma velocities indicative of the low resolution of the spectrograph and suggesting there are systematic errors that have not been accounted for.  

\section{XSHOOTER spectra}

We also observed EPIC212235321 with the medium resolution echelle spectrograph XSHOOTER
\citep{vernet11}, mounted on VLT-UT2 at Paranal, Chile. X-shooter covers the spectral range from the atmospheric cut-off in the UV to the
near-infrared K band in three separate arms, known as the
UVB (0.30 -- 0.56~microns), VIS (0.56 -- 1.01~microns) and NIR
(1.01 -- 2.40~microns).  We observed with XSHOOTER on the VLT on the night of 31 August 2017 and obtained 10 spectra in the UVB and VIS arms with exposure times of 330s each. The data were taken in stare mode, and so the NIR arm data is dominated by the sky background and is not used.  The data were reduced  using the standard pipeline release of the
XSHOOTER Common Pipeline Library (CPL) recipes (version
2.6.8) within ESORex, the ESO Recipe Execution Tool.  The accuracy of the wavelength calibration of XSHOOTER
data from the pipeline reduction is 0.03~nm in the UVB and
0.02~nm in the VIS  arm, corresponding
to a velocity precision of $\sim$10~km~s$^{-1}$ at H$\alpha$. The data were analysed in the same manner as  the SOAR data, although the data were of better quality, and so four Gaussians were used, two for the emission and two for the absorption.  The fit parameters can be seen in Table \ref{rv_tab}. The XSHOOTER data have a resolution R$\sim$7500, higher than the R$\sim$5000 for the SOAR data, hence the errors are much smaller. These are the parameters that are therefore used within the rest of this paper.  We tested the discrepancy between the $\gamma$ values in the XSHOOTER and SOAR data by fixing them to the XSHOOTER values and refitting the SOAR data. The fits then agree to within 1 per cent, indicating our assumption about the underestimated errors on the SOAR data was correct.

\begin{figure}
	\begin{center}
	\includegraphics[width=\columnwidth]{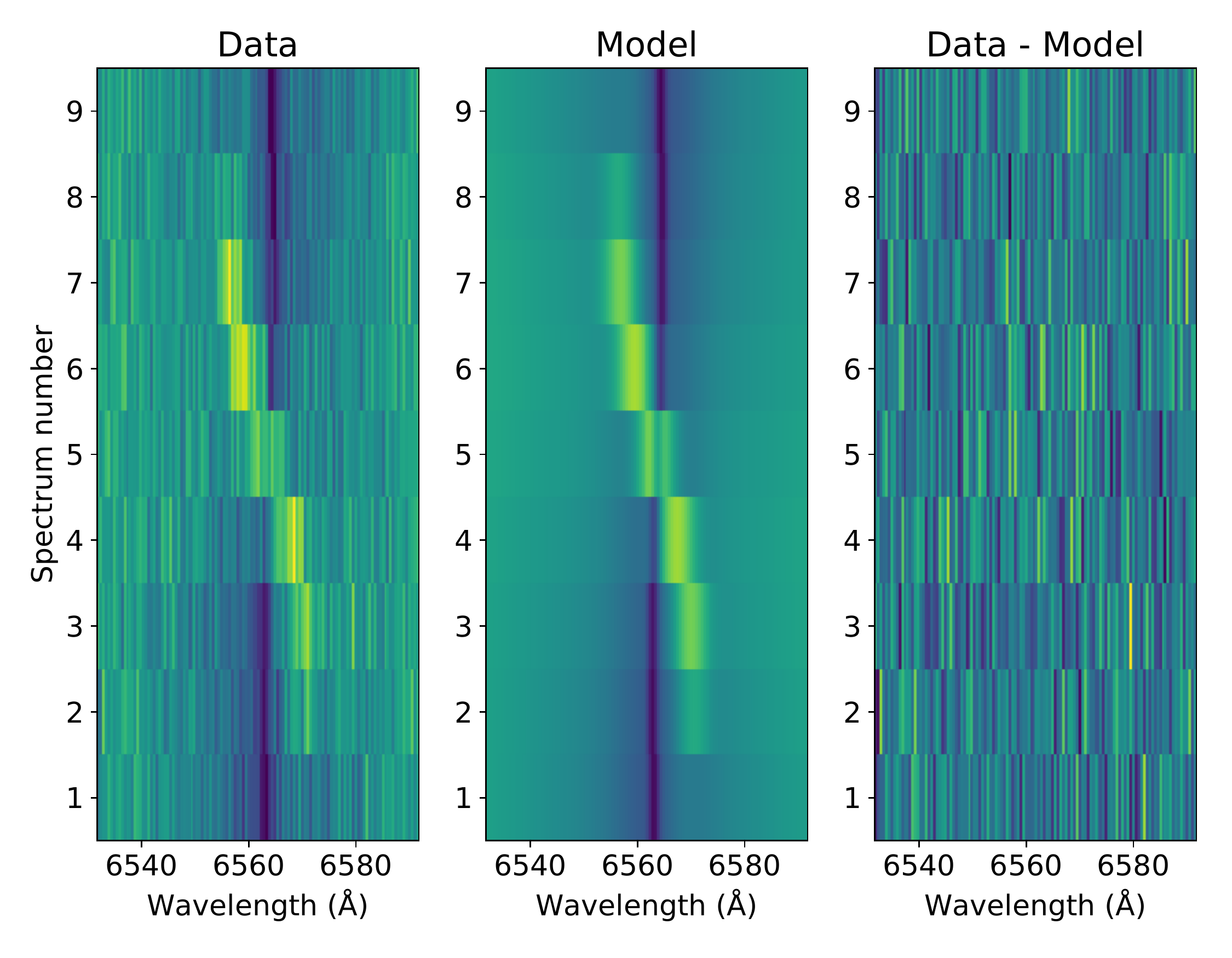}
    \caption{Trailed XSHOOTER spectra showing H$\alpha$ (left), the model fit (centre) and the residuals (right).}
    \label{trail}
\end{center}
\end{figure}

\begin{table}
	\centering
	\caption{\textbf{Derived system parameters, where 1 refers to the white dwarf and em the emission from the brown dwarf.}}
	\label{rv_tab}
	\begin{tabular}{ccc} 
	Property&Value&Units\\ 
    \hline
    K2 P&0.047369569$\pm$0.000000056 & days\\
    K2 T$_0$&2457011.653656$\pm$0.000013& HJD\\
    T$_{\rm eff1}$&24490$\pm$194& K\\
    log g$_{1}$&7.63$\pm$0.02& -\\
    K$_{\rm 1}$&41$\pm$2& kms$^{-1}$\\
    $\gamma_{\rm 1}$&47.5$\pm$1.6& kms$^{-1}$\\
    R$_{1}$&0.017$\pm$0.005&R$_{\odot}$\\
    K$_{\rm em}$&308$\pm$5& kms$^{-1}$\\
    $\gamma_{\rm em}$&35$\pm$5& kms$^{-1}$\\
    R$_{\rm em}$&0.0973&R$_{\odot}$\\
    a&0.44&R$_{\odot}$\\
	\hline

\end{tabular}
\end{table}

The XSHOOTER spectra show evidence of  the Ca~{\sc ii} triplet at 8498.02 8542.09 8662.14 \AA \, and Mg~{\sc i} at 8806.76 \AA \,  in emission within our spectra. We did not detect any Na~{\sc i} in emission, but we do detect interstellar Na~{\sc i} in absorption at 5889.95 and 5895.92 \AA. We also do not detect Fe~{\sc i}, Fe~{\sc ii}, Ti~{\sc i} or K~{\sc i}  in emission, as are seen in WD0137-349B \citep{longstaff17}. There are also no Mg~{\sc ii} or Ca~{\sc ii} absorption features which could be attributed to accretion from a dust disk as are seen in SDSS1557 \citep{farihi17}. SDSS1557  has a similar effective temperature to EPIC212235321, however it only shows Balmer line emission features from the brown dwarf, and not emission from other atoms within the brown dwarf atmosphere.

\section{Discussion}

The  log~g  and T$_{\rm eff}$ of the white dwarf  mean that it falls in a mass range between two types of white dwarf models; the He core and C/O core models. Using the C/O core models of \citet{fonaine01}, the white  dwarf mass is 0.47$\pm$0.01 M$_{\odot}$.  The He core models of \citet{panei07, panei00} give masses of 0.46 M$_{\odot}$ and 0.48 M$_{\odot}$ depending  on the model grid used. These He core grids are quite  coarsely spaced, and the uncertainties due to T$_{\rm eff}$ and log g  are much smaller  than the difference in mass determined between the grids.  As both sets of models give consistent results we take the mass of the white dwarf  to be 0.47 $\pm$0.01~ M$_{\odot}$. The radius determined from the \citet{fonaine01} models is 0.017$\pm$0.005 ~R$_{\odot}$.  

The difference in the $\gamma$ velocities is 13$\pm$6~km~s$^{-1}$. This value is consistent with the gravitational redshift of the white dwarf calculated using the mass and radius which is 16.4$\pm$0.5~km~s$^{-1}$, serving as an independent check.

The mass ratio of the binary taken from the $K$ velocities is $q=M_2/M_1$ = 0.135. Assuming a white dwarf mass of 0.47~M$_{\odot}$ gives a companion mass of 0.063~M$_{\odot}$, or 66.4~M$_{\rm Jup}$. However,  $K_{\rm em}$ only represents the emission from the heated hemisphere of the brown dwarf, not the true centre of mass velocity of the secondary but a lower limit, and hence an upper limit on the mass determination.  This means that the secondary is definitely a brown dwarf ($M<70$ M$_{\rm Jup}$) and that a correction must be applied to get the true centre of mass velocity. We  used the formula from \citet{parsons12} with   $f$=0.5  corresponding to optically thick emission as H$\alpha$ emission from similar systems has been found to be optically thick (e.g. \citealt{parsons17}, \citealt{parsons12}).  Since  EPIC212235321 is not an eclipsing system we have no direct measurement of the radius of the brown dwarf, needed to calculate this correction. We can however, estimate the radius using  the knowledge that the system is not interacting via Roche lobe overflow.

\begin{figure}
	\begin{center}
	\includegraphics[scale=0.5]{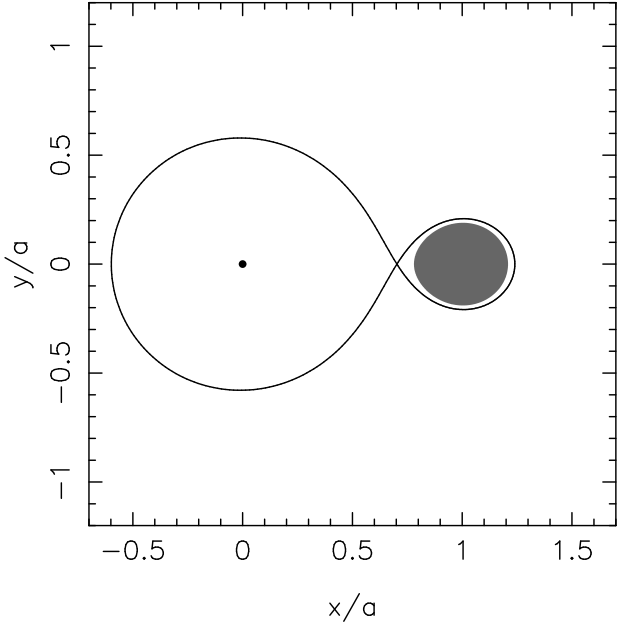}
    \caption{\textbf{Diagram showing the Roche lobes of the white dwarf (left) and brown dwarf (right) in the system for a case with uniform emission (K$_{2}$=346 kms$^{-1}$), and mass ratio of 0.12.. The brown dwarf radius is 0.0973 R$_{\odot}$ and the orbital separation is 0.44 R$_{\odot}$. The white dwarf is not shown to scale.}}
    \label{rlobe}
\end{center}
\end{figure}

We used the  Roche lobe calculations in \citet{breedt12}, and white dwarf  mass and orbital separation, to calculate the Roche lobe for a range of secondary masses. We then compared these predictions on the size of the Roche lobe with the upper limit on the mass determination 66.4~M$_{\rm Jup}$, and radii predicted by the BT-Cond models \citep{allard12}. The size of the Roche lobe is 0.0988$\pm$0.007~R$_{\odot}$, and as there are no signs of accretion within this system, this is the maximum size possible for the secondary . Using this brown dwarf radius allows us to determine that the inclination of the system is likely to be greater than 48 degrees, which when combined with the K correction, gives a mass range for the secondary of $47<$M$_2<65$ M$_{\rm Jup}$. If we assume the emission feature is due to uniform emission from the heated hemisphere of the brown dwarf, then the inclination of the system is 56 degrees, and the brown dwarf mass is 58 M$_{\rm Jup}$. In the following discussions, this is taken to be the mass of the brown dwarf. Figure \ref{rlobe} is a diagram of the system showing both binary constituents and their Roche lobes.

The BT-COND evolutionary models of brown dwarfs indicate that in order for the brown dwarf to not fill its Roche lobe at this orbital period, it must be older than 700~Myr with a radius of 0.0973 R$_{\odot}$. Such a brown dwarf has T$_{\rm eff} \sim$2000 K, which gives an estimated spectral type of early L, around L3 \citep{golimowski04}. If the brown dwarf is older, the effective temperature is likely to be cooler. The cooling age of the white dwarf was estimated using the white dwarf models, resulting in  18 $\pm$1~Myr from the C/O core models and  14$\pm$8~Myr from the He core models. Clearly, while the white dwarf is relatively newly formed, the system itself is much older.

While we cannot use an initial-final mass relationship for white dwarfs (e.g. \citealt{casewell09}) to determine the mass of the white dwarf progenitor, it is possible to use main sequence models to determine the maximum progenitor mass. We have ignored the cooling time and the time for the common envelope phase as they are so short in comparison to the lifetime of the system (see \citealt{parsons17} for a more detailed description of the common envelope phase), and have assumed that the total system age is equal to the main sequence lifetime.  The \citet{marigo17} PARSEC-COLIBRI models predict that at 700~Myr, the minimum age permitted, the maximum progenitor mass is 2.60~M$_{\odot}$, at 4~Gyr, the maximum progenitor mass is 1.35~M$_{\odot}$, while if the system is older, say 10~Gyr, this drops to 1.04~M$_{\odot}$. The initial-final mass relation for white dwarfs (e.g. \citealt{casewell09}) predicts that a 1 M$_{\odot}$ star should produce a white dwarf with mass near to the average white dwarf mass of $\sim$0.6 M$_{\odot}$, indicating that the evolution of the white dwarf progenitor has been truncated by the common envelope phase.

The period of this system is very short, $\sim$ 3 minutes shorter than that of the two newly discovered  K2 systems presented in \citet{parsons17}, and the masses of the components are very similar. The evolution of these systems was also discussed in depth  in \citet{rappaport17}, and so for further discussion of how these systems will evolve into a CV, we refer to these two papers.

\begin{figure}
 \begin{center}
 \includegraphics[width=\columnwidth]{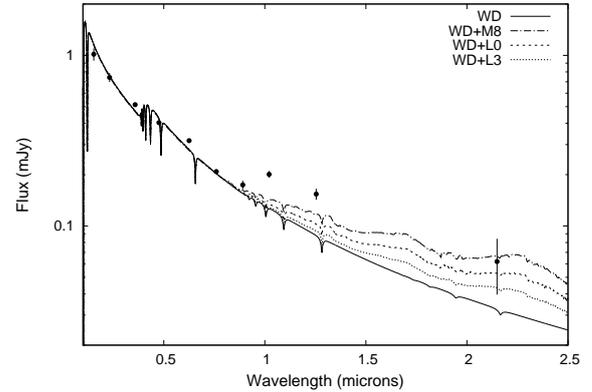}
    \caption{Photometry of EPIC212235321 from \textit{Galex}, VST ATLAS and the VISTA VHS, with a T$_{\rm eff}$ = 24490$\pm$150 K, log g = 7.63$\pm$0.02 white dwarf model, and template M and brown dwarf spectra. The 3$\sigma$ error bars are shown.}
     \label{fig:sed}
     \end{center}
 \end{figure}

Although this system is not eclipsing, and so we have no direct measurement of the radius of the brown dwarf, we were able to independently investigate the possible spectral type of the secondary by using SED fitting. EPIC212235321 is in both the VST ATLAS DR3 \citep{shanks15} and VISTA VHS DR4 catalogues providing $ugrizYJHK_s$ photometry. The VISTA Data Flow System pipeline processing and science archive are described in \citet{irwin04}, \citet{hambly08} and \citet{cross12}.  We converted the VST ATLAS AB magnitudes, the VISTA Hemisphere Survey Vega magnitudes, and the \textit{Galex} AB magnitudes shown in Table \ref{tab:info} into flux in mJy using the relevant zeropoints. We also adjusted the \textit{Galex} magnitudes using the non-linearity transformations presented in \citet{camarota14}. These fluxes were then compared to a white dwarf model of the primary star, as well as combined white dwarf-brown dwarf/M dwarf templates of spectral type M8 (VB10; \citealt{burgasser04}), L0 (2MASP J0345432+254023; \citealt{burgasser06}) and L3 (2MASSW J1146345+223053; \citealt{burgasser10}).  Figure \ref{fig:sed} shows that there is clearly an excess in the near-IR wavelengths. However, the excess is indicative of spectral type earlier than M8 in the $Y$ and $J$ bands, and allows as late a spectral type of L3 in the $K_s$ bands. At the estimated age of the brown dwarf, the masses 
indicate the effective temperatures should be $\sim$ 2000 K depending on the mass of the secondary and the age. This effective temperature is consistent with a spectral type of L3 \citep{golimowski04}.

We used the \citet{jura03} disk model to determine if it is possible some of the excess seen is due to a circumbinary dust disk as for J155720.77+091624.6 \citep{farihi17}. We used the parameters given in \citet{debes11} for the outer an inner disk radius. These parameters, when combined with the white dwarf parameters give an outer disk temperature of 600 K, and an inner temperature of $\sim$1500 K, close to the sublimation temperature. Such a disk however, only begins to make a noticeable contribution within the SED of the combined system in the $K_s$ band - the flux is negligible in $YJ$. When we combine this with the fact that we see no pollution in the XSHOOTER spectrum indicative of the white dwarf being a DAZ, we can discount the presence of a dust disk in the system.

It is possible  that the majority of the "excess" seen is due to a large reflection effect. The VISTA data were taken at phases 0.46 in $Y$, 0.55 in $J$ and 0.65 in $K_s$, meaning that the $Y$ photometry was observed as the hot side of the brown dwarf comes into view, and the $J$ band was observed just past the peak emission from the brown dwarf at phase 0.5.  In order for the excess seen to be due to the reflection effect in the system, we would have to have variations with semi-amplitudes of $\sim$0.25-0.3 mags in the $Y$ and $J$ bands.  For comparison, the lightcurve of WD0137-349AB, varies by $\sim$ 15 per cent in the $K_s$ band. EPIC212235321 shows this level of variability in the $i$ band, due to its much shorter period, and hotter white dwarf. Thus we expect large variability at longer wavelengths (e.g. in the mid-IR) from this system. Indeed, JWST will provide an excellent suite of instruments with which to study these shorter period, highly irradiated brown dwarfs in more detail.

\section{Conclusions}
We have discovered EPIC212235321, the first sub-70 minute non-interacting WD+BD binary known. The lightcurves of this system show a significant reflection effect, which is supported by a large near-IR excess when the VISTA magnitudes are compared to a model white dwarf atmosphere combined with brown dwarf spectra.  There are emission features seen  from the hydrogen  balmer lines as well as Ca II and Mg I, similar to the emission features seen in WD0137-349B \citep{longstaff17}. The radial velocity measurements of the system suggest that the mass of the white dwarf is 0.47 $\pm$0.01~ M$_{\odot}$, and the mass of the brown dwarf companion is 58 M$_{\rm Jup}$, consistent with a T$_{\rm eff}$=2000 K and a spectral type of L3. This system, along with SDSS J1205-0242,  and SDSS J1231+0041 \citep{parsons17} are the shortest period detached post-common envelope binaries containing a white dwarf and a brown dwarf known. All three have periods shorter than those of some CVs that are actively accreting (e.g.\citealt{hern16, burleigh1212}). These three systems also have low mass white dwarf primaries suggesting common envelope evolution has truncated the lifetime of the red giant star that was the white dwarf progenitor.

\section{Acknowledgements}
S.L. Casewell acknowledges support from the University of Leicester Institute for Space and Earth Observation and I.P Braker acknowledges support from the University of Leicester College of Science and Engineering. S.G. Parsons acknowledges the support of the Leverhulme Trust. Support for this work was provided by NASA through Hubble Fellowship grant HST-HF2-51357.001-A, awarded by the Space Telescope Science Institute, which is operated by the Association of Universities for Research in Astronomy, Incorporated, under NASA contract NAS5- 26555.  This work is based on observations obtained at the Southern Astrophysical Research (SOAR) telescope, which is a joint project of the Minist\'{e}rio da Ci\^{e}ncia, Tecnologia, Inova\c{c}\~{a}os e Comunica\c{c}\~{a}oes (MCTIC) do Brasil, the U.S. National Optical Astronomy Observatory (NOAO), the University of North Carolina at Chapel Hill (UNC), and Michigan State University (MSU). This paper also uses observations made at the South African Astronomical Observatory (SAAO). The WHT is operated on the island of La Palma by the Isaac Newton Group of Telescopes in the Spanish Observatorio del Roque de los Muchachos of the Instituto de Astrof\'{i}sica de Canarias. Also based on data products from observations made with ESO Telescopes at the La Silla Paranal Observatory under program ID 177.A-3011(A,B,C,D,E,F,G,H,I,J), and 099.D-0252.
\bibliographystyle{mnras}

\bibliography{bib}

\label{lastpage}
\end{document}